\documentclass[12pt,a4paper,english,fleqn]{article}
\usepackage[sort&compress,round,comma,authoryear]{natbib}
\usepackage[T1]{fontenc}
\usepackage{ae,aecompl}
\usepackage[colorlinks=true,urlcolor=blue,citecolor=blue,linkcolor=red,bookmarks=true]{hyperref}
\usepackage{graphicx}
\usepackage{float}	% Including figure files
\usepackage{amsmath}	% Advanced maths commands
\usepackage{amssymb}	% Extra maths symbols
\usepackage{amsfonts}
\usepackage{babel}
\usepackage{soul}
\newlength{\lp}
\makeatother
\begin{document}
\title{The origin of the MOND critical acceleration scale.}
\author{D. F. Roscoe \\ Dept of Mathematics \\The Open University \\ Milton Keynes MK7 6AA \\ D.Roscoe@open.ac.uk\\ \\ORCID: 0000-0003-3561-7425}
\date{}
\maketitle
\newpage
\begin{abstract}
The irrefutable successes of MOND are predicated upon the idea that a critical gravitational acceleration scale, $a_0$, exists. But, beyond its role in MOND, the question \emph{Why should a critical gravitational acceleration scale exist at all?} remains unanswered. There is no deep understanding about what is going on. 
\\\\
Over roughly the same period that MOND has been a topic of controversy, Baryshev, Sylos Labini, Pietronero and others have been arguing, with equal controversy in earlier years, that, on medium scales at least, galaxies are distributed in a $D\approx 2$ quasi-fractal fashion. There is a link: if the idea of a $D\approx 2$ quasi-fractal universe on medium scales is taken seriously then there is an associated mass surface density scale, $\Sigma_F$ say, and an associated Newtonian gravitational acceleration scale, $a_F \equiv 4\pi G \,\Sigma_F$. If, furthermore, the inter-galactic medium (IGM) exhibits the same quasi-fractal structure then it is an obvious step to consider the possibility that $a_0$ and $a_F$ are one and the same thing.
\\\\
Subsequently, via a modern geometric realization of the Leibniz-Mach worldview, we obtain a detailed theoretical understanding of how galaxy disks should interact with a $D\approx 2$ quasi-fractal IGM.
This understanding takes the form of a superficially unremarkable scaling relationship which, used with standard photometric mass-modelling applied to SPARC data, shows that $a_F \approx 1.2\times10^{-10}\,mtrs/sec^2$ is explicitly embedded in that data. Since the scaling relationship also gives rise to the Baryonic Tully-Fisher Relationship, but with $a_0$ replaced by $a_F$, we are led unambiguously to the conclusion that $a_0$ and $a_F$ are, in reality, one and the same thing. 
\end{abstract}
\newpage
%\begin{keywords}
%{\bf Key words:} Galaxies: general; gravitation
%\end{keywords}

%%%%%%%%%%%%%%%%%%%%%%%%%%%%%%%%%%%%%%%%%%%%%%%%%%

%%%%%%%%%%%%%%%%% BODY OF PAPER %%%%%%%%%%%%%%%%%%
\section{Introduction:}\label{Intro}
Over the years since Milgrom conceived MOND (\citet{Milgrom1983a, Milgrom1983b,Milgrom1983c, Milgrom1983d, Milgrom1983e}), significant effort has been expended on devizing various variational principles designed to reproduce its details. The very first of these to appear was AQUAL (\citet{Milgrom1984}) although there are now several competing flavours of such derivative theories all of which seek, one way or another, to make the fundamentals of MOND compatible with those of General Relativity. 
\\\\
However, the recent paper of \citet{Chae} reports a high-confidence detection of the External Field Effects (EFEs) predicted by MOND to affect galaxy discs in dense environments - crucially, the authors report that the detection of these EFEs points to a breakdown of the Strong Equivalence Principle (SEP). Consequently, whilst something like MOND is supported, the fundamental principle supporting General Relativity is undermined - the two worldviews cannot be reconciled. 
\\\\
But if something like MOND is supported, then we must seek to make deeper sense of the ideas which underly it. To this end, it is striking that in all of the theoretical output around these ideas there has been virtually no discussion about the significance of Milgrom's primary insight - the existence of a critical gravitational acceleration scale, $a_0$ say, in the first place. Instead, $a_0$ is routinely consigned to the passive role of simply signposting the boundary where one gravitational regime gives way to another.  
\\\\
But if we are ever to come to a fundamental understanding of the mechanisms which MOND emulates to some degree, then  the question \emph{Why should a critical gravitational acceleration scale exist at all?} must necessarily be posed and answered. 
It is precisely this question which connects Milgrom's work to that of  \citet{Baryshev1995}  and many others (see appendix \S\ref{Observations} for an overview)  which has shown how the distribution of galaxies on the medium scale ($\approx$ 100Mpc at least) is quasi-fractal, $D\approx 2$.  The connection is simply this: if the idea of a quasi-fractal universe on medium scales is taken seriously to the extent that the distribution of material within the IGM itself has the same quasi-fractal quality then, within the IGM, there is an associated mass surface density scale, $\Sigma_F$ say, and an associated Newtonian gravitational acceleration scale, $a_F = 4\pi G \,\Sigma_F$. It is then a natural step to consider the possibility that $a_0$ of MOND and $a_F$ of the postulated $D\approx 2$ quasi-fractal IGM are one and the same thing.
\\\\
Regardless of the precise material properties of this postulated IGM, a quantitative model with the potential to consider the possibility that $a_0 \equiv a_F$ already exists in the mainstream literature in an early form  as \citet{Roscoe2002} and in mature form as \citet{Roscoe2020} in the arXiv. This model, which is a modern geometric realization of the Leibniz-Mach worldview (appendix \S\ref{LMC}), has an irreducible equilibrium state (appendix \S\ref{ESLC}) in which universal material necessarily exists as a non-trivial fractal $D = 2$ distribution, thereby providing the theoretical link to the work of  Baryshev, Sylos Labini, Pietronero et al and hence, potentially, to that of Milgrom.
\\\\
Interpreting this irreducible equilibrium state as an idealized representation of the postulated $D\approx 2$ quasi-fractal IGM, and noting that
the idea of  a critical acceleration boundary around a galactic object amounts to the idea that the galactic object concerned has a finite boundary around it, then the way forward is easily recognized: a galactic object is simply modelled as a finite bounded perturbation of the irreducible equilibrium state. We develop the model for the simplest possible case in which the perturbation is spherical and the perturbation boundary, $R=R_0>0$ say, is finite but initially undetermined.
\\\\
We then consider the special case which refers to purely circular motions from which the superficially unremarkable scaling relation:
\begin{equation}
\left[ \frac{V_{rot}(R)}{V_{flat}} \right]^2 \Sigma_R = \Sigma_F;~~~R < \infty \label{FirstEqn}
\end{equation}
emerges,
where $\Sigma_R \equiv \Sigma(R, R_0, M_0)$ represents the mass surface density of material around the galactic object at radius $R$, and $M_0$ is the mass contained within the perturbation boundary $R_0$. In passing, we can note that Renzo's Rule follows directly from the structure of this scaling relation.
\\\\
More generally, according to (\ref{FirstEqn}), the dynamics within any given galaxy are tightly constrained by an External Matter Field - the fractal IGM characterized by  $\Sigma_F$. In the terminology of MOND, this means that an External Field Effect (EFE) is the dominating control which determines disk dynamics and implies that knowledge of $\Sigma_F$ (and hence of $a_F$) is explicitly embedded in the internal structures of disks and disk dynamics. Consequently, if $a_0\equiv a_F$ then we have the prediction that $a_F \approx 1.2\times10^{-10}\,mtrs/sec^2$ is explicitly embedded in SPARC rotation curves and photometry.
\\\\
In practice, when (\ref{FirstEqn}) is supplemented by a condition of stable equilibrium on the perturbation boundary then there are six primary consquences: 
\begin{itemize}
\item The EFE manifests itself through two particular symmetries which leave (\ref{FirstEqn}) invariant, and one of these provides for the direct evaluation of  $a_F$, and hence of $\Sigma_F$, from rotation curves \& photometry.
Applied to the SPARC sample of \citet{Lelli2016A} with a fixed stellar MLR, $\Upsilon_* \in(0.5,\, 1.0)$, and using the mass models of \citet{Lelli2016B}, this symmetry gives $a_F \approx 1.2\times10^{-10}\,mtrs/sec^2$;
\item The Baryonic Tully-Fisher Relation (BTFR) emerges automatically, but with $a_0$ replaced by $a_F\equiv 4\pi G\Sigma_F$. This incorporation of $\Sigma_F$ into the BTFR implies that the BTFR itself is a manifestation of an EFE;
\item The conclusion that $a_0 \equiv a_F$ follows directly from the two results above;
\item The process provides a dynamical means for estimating absolute radial scales for disks and hence their absolute distance scales;
\item MOND is seen to work as  successfully as it does because, in practice, the far-field form of the MOND force-law is actually emulating the $\Sigma_F$ External Matter Field;
\item The net result of these considerations is that, in practice, Milgrom's MOND is shown to be  firmly anchored in the geometrical model of a Leibniz-Mach worldview.
\end{itemize}
In the following, the working hypothesis that $a_0 \equiv a_F$ is motivated via a primitive model in \S\ref{SimpleModel}, the core analysis is given in \S\ref{OverView} (and specifically pivots on the results of \S\ref{EFEs} \& \S\ref{aFcalc}) whilst the implications for the IGM are considered in \S\ref{IGM}.
\section{A primitive Newtonian model}\label{SimpleModel}
We use a primitive Newtonian model to motivate the  working hypothesis that $a_0 \equiv a_F$.
\\\\
As we have noted, the work of \citet{Baryshev1995}  and many others (see appendix \S\ref{Observations} for an overview)  has shown how the distribution of galaxies on the medium scale ($\approx$ 100Mpc at least) is quasi-fractal, $D\approx 2$.  If the idea of a quasi-fractal universe on medium scales is idealized to include all physical scales then  we can say that, \emph{about any point} chosen as the centre, mass is distributed according to
\begin{equation}
\mathcal{M}(R) = 4 \pi R^2 \,\Sigma_F; \,\,\,R < \infty, \label{Frac1}
\end{equation}
where $\Sigma_F$ is the characteristic mass surface density of the distribution. 
\\\\
Since this idealized matter distribution is isotropic (by definition) about any arbitrarily chosen centre, then the notional Newtonian gravitational acceleration imparted to a particle at radius $R$, and generated by the material contained within $R$, is directed towards the chosen centre and has magnitude given by
\[
a_F \equiv \frac{\mathcal{M}(R) \, G}{R^2} = 4 \pi G\, \Sigma_F;\,\,\, R< \infty.
\]
On this basis, we note that: 
\begin{itemize}
	\item The \emph{net} actual gravitational acceleration imparted to a material particle immersed in the global distribution (\ref{Frac1}) is zero;
	\item If a finite spherical volume, radius $R_0$, is imagined emptied of all material, then the net actual gravitational acceleration of any material particle placed on $R_0$  will be $a_F$ directed radially outwards from the centre of the empty volume;
	\item A material particle placed on $R < R_0$ will experience an outward acceleration $a_F^* < a_F$ which satisfies $a_F^* = 0$ when $R=0$, and which is a primitive form of an external field effect;
	\item A material particle placed on $R > R_0$ will experience an outward acceleration $a_F^*< a_F$ which is such that $a_F^* \rightarrow 0$ as $R \rightarrow \infty$;
	\item The empty spherical volume is unstable since all accelerations on $R_0$ are outward. It follows that stability requires the volume to be occupied by a stablizing mass, a galaxy say, creating a state of zero net radial acceleration on $R_0$.
\end{itemize}
{\bf Galaxy Equilibrium Condition:} Zero net radial acceleration on $R_0$ requires
\begin{equation}
g_0 = a_F \label{eqn2HH}
\end{equation}
where $g_0$ is the gravitational acceleration generated on $R_0$ by the contained galaxy.
\\\\
Given that the statement $g_0 = a_0$ is fundamental to MOND, we take (\ref{eqn2HH}) to justify the working hypothesis that $a_0 \equiv a_F$.
Whilst the foregoing primitive model has no power to predict or quantify anything (beyond global equilibrium), it has the benefit of providing a de-mystified interpretation of $a_0$ and creating a rational basis upon which this interpretation can be explored. 
\section{Galaxy dynamics in the $\Sigma_F$ environment} \label{OverView}
The rational account of the MOND critical acceleration scale arising from the working hypothesis $a_0 \equiv a_F$, and developed in the following, has its source in a modern geometric realization of the Leibniz-Mach worldview. An outline is given in \S\ref{LMC}.
We use this to derive the most simple possible galaxy model (spherically symmetric with circular motions only), which presents as the superficially unremarkable scaling relationship (\ref{eqn2}) below. 
\\\\
A detailed understanding of the workings of External Field Effects (EFEs) in this simple case emerges directly from this scaling relationship, and is described in \S\ref{EFEs} \& \S\ref{aFcalc}. These results lead immediately to the evaluation of $a_F$ in \S\ref{Gamma1} where it is found that $a_F\approx 1.2\times 10^{-10}\,mtrs/sec^2$. Consequently, the BTFR (but with $a_0$ replaced with $a_F$) is derived in \S\ref{BTFR}. Taken together, these results lead to the unambiguous conclusion that $a_0\equiv a_F$.
\subsection{The galaxy model and the IGM:}
The irreducible basic expression of the global geometric Leibniz-Mach model is a world of dynamic equilibrium within which material is necessarily non-trivially present in the form of a $D=2$ fractal distribution so that, about any point chosen as the centre, mass is distributed according to (\ref{Frac1}). We assume this to be a fair description of the unperturbed IGM.
This latter assumption raises obvious questions concerning the nature of the material populating the IGM which are discussed in detail in \S\ref{IGM}. 
\\\\
Given the foregoing, and noting that the idea of a critical acceleration boundary around a galactic object amounts to the idea that the galactic object concerned has a finite boundary around it, then an individual galaxy is modelled as a finite bounded spherical perturbation of (\ref{Frac1}) so that it has the general form:
\begin{eqnarray}
\mathcal{M}(R) &\equiv& \mathcal{M}_g(R),~~~~ R \leq R_0; \label{Frac2} \\
\mathcal{M}(R) &\equiv&  \mathcal{M}_g(R_0)  + 4 \pi (R^2-R_0^2)\,\Sigma_F,~~~~ R > R_0, \nonumber
\end{eqnarray}
where $\mathcal{M}_g(R)$ represents the galactic mass contained within radius $R\leq R_0$ and $R_0$ is the finite, but otherwise unspecified, radial boundary of the perturbation. 
\\\\
For later clarity, we write
\begin{equation}
\mathcal{M}_g(R) \equiv M_0 \,F(R/R_0);~~ F(0) = 0,~F(1) = 1,~~~R \leq R_0, \label{Frac3}
\end{equation}
where $F(R/R_0)$ describes the relative distribution of mass in $R\leq R_0$ and $M_0$ specifies the total mass contained within $R=R_0$.
\\\\
For the special case of purely circular orbits the equations of motion deriving from (\ref{4F}) integrate to give (\ref{3B}). This, with mass model (\ref{Frac2}), notation (\ref{Frac3}) and writing $v_0 \equiv V_{flat}$ becomes  the scaling relationship:
\begin{eqnarray}
\left[\frac{V_{rot}(R)}{V_{flat}}\right]^2 \Sigma_R &=& \Sigma_F, ~~~~ R < \infty; \label{eqn2} \\  
\Sigma_R &\equiv& \frac{M_0\,F(R/R_0)}{4 \pi R^2},~~~~ R \leq R_0; \nonumber \\
\Sigma_R &\equiv& \frac{M_0  + 4 \pi (R^2-R_0^2)\,\Sigma_F}{4 \pi R^2},~~~~ R > R_0 \nonumber
\end{eqnarray}
where $M_0$ is the mass contained within the perturbation boundary $R = R_0$. This is the expanded form of (\ref{FirstEqn}).
\subsection{Symmetries \& EFEs}\label{EFEs}
The scaling relationship (\ref{eqn2}) is invariant under three distinct symmetry transformations each of which generates a one-parameter class of solutions to (\ref{eqn2}):
\begin{equation}
\Sigma_F \rightarrow \alpha \, \Sigma_F,~~~~~~ R \rightarrow \alpha^{-1/2}\, R; \label{eqn2B}
\end{equation}
\begin{equation}
M_0 \rightarrow \beta \, M_0,~~~~~~ R \rightarrow \beta^{1/2}\, R; \label{eqn2A}
\end{equation}
\begin{equation}
\Sigma_F \rightarrow \gamma \,\Sigma_F,~~~~~~ M_0 \rightarrow \gamma\, M_0. \label{eqn2C}
\end{equation}
Of these, (\ref{eqn2B}) provides for a direct theoretical evaluation of $\Sigma_F$ (and hence of $a_F$), whilst (\ref{eqn2A}) is essential for its practical evaluation from SPARC rotation curves \& photometry.  Symmetries (\ref{eqn2B}) and (\ref{eqn2C}) together unambiguously demonstrate the deep connection between the $\Sigma_F$ External Matter Field and the internal properties of the galactic object. They are the formal representation   of EFEs in disk structure and dynamics for an isolated galaxy immersed in the External Matter Field.
\subsection{Theoretical evaluation of $\Sigma_F$} \label{aFcalc}
Using $a_F = 4 \pi G \Sigma_F$ and identifying the gravitational acceleration $g_0$ with the centripetal acceleration, then (\ref{eqn2B}) gives:
\begin{equation}
a_F \rightarrow \alpha\, a_F;~~~~g_0 \rightarrow \alpha^{1/2}\,g_0 ~~{\rm where}~~g_0 \equiv \frac{V_0^2}{R_0}
\label{eqn2J}
\end{equation}
from which it is immediate that
\begin{equation}
\Gamma \equiv \frac{g_0^2}{a_F}, \label{eqn2BB}
\end{equation}
which has dimensions of $mtrs/sec^2$, is an invariant of (\ref{eqn2}) under the symmetry (\ref{eqn2B}).  In other words, the value of $\Gamma$ is independent of any value assigned to $\Sigma_F$, and hence to $a_F$ - and this has the direct consequence that an \emph{arbitrary} positive value for $\Sigma_F$ in (\ref{eqn2}) can be used in the curve fitting process without affecting the computed value for $\Gamma$.
\\\\
Furthermore, if we now apply the Galaxy Equilibrium Condition of (\ref{eqn2HH})  so that
\begin{equation}
g_0 = a_F, \label{eqn2CC}
\end{equation}
then, in practice, we are selecting a particular rescaling of $(\Sigma_F,\,R)$ under symmetry (\ref{eqn2B}). Applied to (\ref{eqn2BB}), this rescaling gives 
\begin{equation}
g_0 = a_F = \Gamma,   \label{eqn2DD}
\end{equation}
so that $\Sigma_F$ is also determined. Finally, from this latter result, we have directly
\begin{equation}
R_0 = \frac{V_0^2}{\Gamma} \label{eqn2EE}
\end{equation} 
for a dynamical determination of the absolute radius of the perturbation boundary.
\subsection{Practical evaluation of $(a_F,\,\Sigma_F)$} \label{Gamma1}
\begin{figure}[H]
	\centering
	\includegraphics[width=0.7\linewidth]{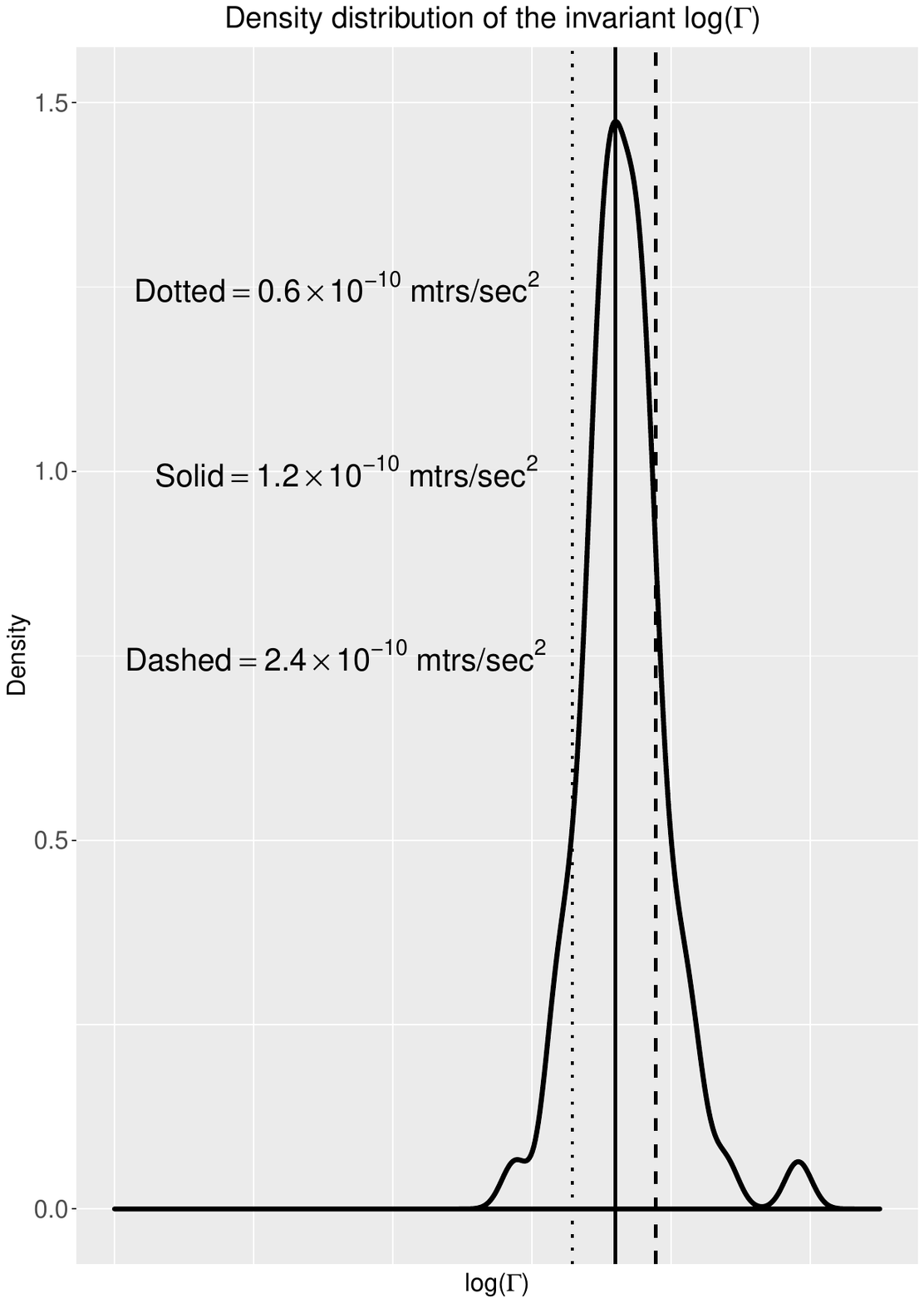}
	\caption{Solid black curve = density distribution of $\log(\Gamma)$. Here, $\Upsilon_* = 0.8$. }
	\label{GammaDensity}
\end{figure}

We summarize the results obtained from the detailed algorithm given in appendix \S\ref{MassModel}. 
\\\\
The basic resource is the SPARC sample of \citet{Lelli2016A}, for which the object selection criteria is described in \S\ref{Gamma}.
Following \citet{Lelli2016B}, all required baryonic mass determinations in galaxy disks are then provided by the model
\begin{equation}
M_{bar} = M_{gas} + \Upsilon_*\, L_{[3.6]}, \label{MassModel1}
\end{equation} 
where $M_{gas}$ is the gas mass, $L_{[3.6]}$ is the $[3.6]$ luminosity and $\Upsilon_*$ is the stellar MLR, assumed here to be equal for the disk and the bulge (where present), and constant across the whole SPARC sample.
Lelli et al find that the optimal choice for $\Upsilon_*$ (which minimizes scatter in the BTFR) approximately satisfies $\Upsilon_* \geq 0.5$. For current purposes we restrict the choice to $\Upsilon_* \in (0.5,\,1.0)$.
\\\\
Following the theory of \S\ref{aFcalc}, the primary quantity to be determined in the first instance is the invariant $\Gamma$ of (\ref{eqn2BB}).
In practice, because of uncertainties in estimating the absolute position of the perturbation boundary on the rotation curve then, across the whole sample with a globally applied stellar MLR, $\Upsilon_* \in (0.5,\,1.0)$, we get distributions of values for $\Gamma$ similar to that plotted in figure \ref{GammaDensity} for which $\Upsilon_*=0.8$. These distributions are sharply modal and, using the modal values of these distributions as the estimate for $\Gamma$ in each case, we find:
\[
\Gamma \in (1.4,\,0.9) \times 10^{-10} \, mtrs/sec^2.
\]
Consequently,  applying the Galaxy Equilibrium Condition (\ref{eqn2CC}) to (\ref{eqn2BB}) we get: 
\[
g_0 = a_F = \Gamma \in (1.4,\,0.9) \times 10^{-10}\,mtrs/sec^2. 
\]
In particular, the choice $\Upsilon_* = 0.8$ yields $a_F \approx 1.2 \times 10^{-10}\,mtrs/sec^2$ which corresponds to $\Sigma_F \approx 0.14\, kg/mtrs^2$ for the mass surface density of the External Matter Field;
\subsection{The BTFR} \label{BTFR}
\begin{figure}[H]
	\centering
	\includegraphics[width=0.65\linewidth]{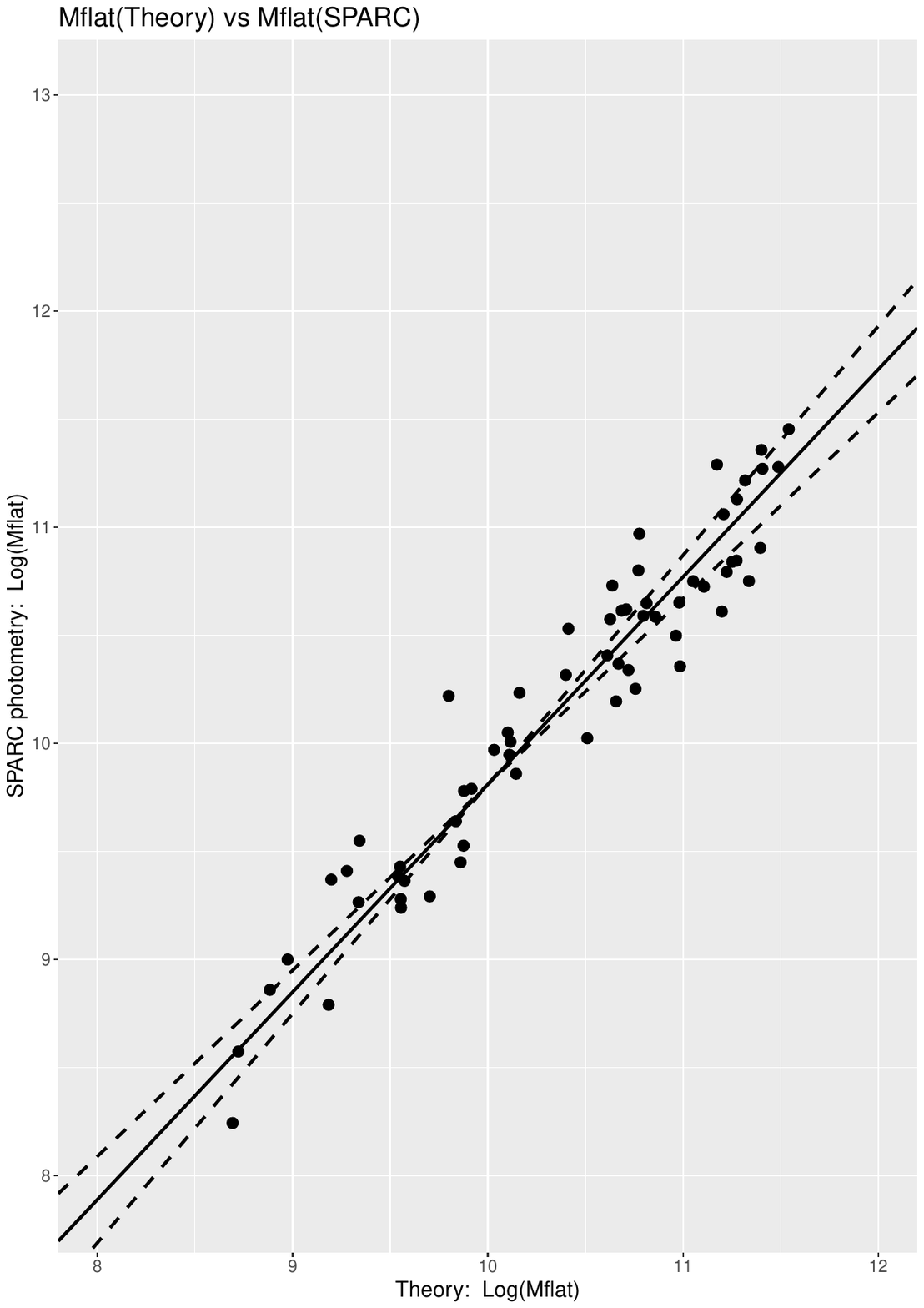}
	\caption{ $M_{flat}$(theory) vs $M_{flat}$(SPARC).}
	\label{fig:SSM-2}
\end{figure}

From (\ref{eqn2}a) we have, on the $R=R_0$ boundary:   
\begin{equation}
V_0^2 = V_{flat}^2 \,\frac{\Sigma_F}{\Sigma_0}=  V_{flat}^2 \left( \frac{ a_F  R_0^2}{G M_0}\right) \label{Frac6}
\end{equation}
where the external field relation $a_F = 4 \pi G \Sigma_F$ has been used. 
Eliminating $R_0$ between the Galaxy Equilibrium Condition  (\ref{eqn2CC}) and (\ref{Frac6}) gives directly:
\begin{equation}
V_{flat}^4 = a_F\, G\, \left[ \left(\frac{V_{flat}}{V_0} \right)^2 M_0 \right]  \label{eqn4d}
\end{equation}
Defining $M_{flat}$(theory) according to the scaling relation
\begin{equation}
M_{flat}{\rm(theory)} \equiv \left(\frac{V_{flat}}{V_0} \right)^2 M_0, \label{eqn5d}
\end{equation}
then (\ref{eqn4d}) becomes
\begin{equation}
V_{flat}^4 = a_F\, G\, M_{flat}{\rm(theory)}  \label{eqn5c}
\end{equation}
which has the exact structure of Milgrom's form of the empirical BTFR, except for the replacement of $a_0$ by $a_F$. 
So, everything hinges on the extent to which $M_{flat}$(theory) tracks $M_{flat}$(photometric), where estimates of the latter are supplied for a significant number of objects in the SPARC database.
\\\\
However, before continuing, there is an issue which must be resolved: specifically, on those RCs for which $V_0 > V_{flat}$ then the rotation velocity $V=V_{flat}$ occurs \emph{twice} on the RC - once on the rising part \emph{before} $V=V_0$, and once asymptotically on the falling part \emph{after} $V=V_0$. Consequently, in such cases, there \emph{appears} to be an ambiguity around what is meant by $M_{flat}$(theory). However since, for such cases, (\ref{eqn5d}) shows that we necessarily must have $M_{flat}$(theory)$ < M_0$ then it is clear that $M_{flat}$(theory) must be defined to mean the mass contained within the \emph{first} occurence of $V = V_{flat}$.
\\\\
With this understanding, the scatter plot given in figure \ref{fig:SSM-2} of $\log{M_{flat}}({\rm photometry})$ against $\log M_{flat}$(theory) makes it clear that these two quantities are in an almost perfect statistical correspondence, and a \emph{least-area} linear regression (a method which treats $x$ and $y$ data identically, see \S\ref{LeastAreas}) gives:
\begin{equation}
\log M_{flat}({\rm photometry}) =   \left(0.96 \pm 0.10 \right) \log M_{flat}({\rm theory}) +  \left(0.21 \pm 1.00 \right)  \label{eqn6d}
\end{equation}
so that, via (\ref{eqn4d}) and (\ref{eqn5d}), the BTFR is implicitly satisfied.
To make the point that the BTFR itself is explicitly satisfied, we find
\begin{equation}
\log M_{flat}({\rm photometry}) =   \left(3.80 \pm 0.58 \right) \log V_{flat} +  \left(1.94 \pm 1.20 \right).  \label{eqn6e}
\end{equation}
\subsection{MOND \& External Field Effects} \label{Discussion}
The foregoing exposition of the dominant role EFEs play in disk dynamics appears to conflict very strongly with the MONDian view which, whilst considering EFEs to be a phenomonological necessity (\citet{Sanders}), views them as subtle and very difficult to detect (see \citet{Chae}). 
\\\\
This apparent conflict is resolved by recognizing that the MOND far-field force law itself works as it does because it is actually emulating the $\Sigma_F$ External Matter Field, giving rise to primary EFEs which are not recognizable as such from within the MONDian context. This interpretation becomes self-evident when we consider how the BTFR arises from each point of view.
\\\\
In the Leibniz-Mach modelling exercise, the BTFR arises directly from the integrated equation of motion (\ref{FirstEqn}) constrained by the equilibrium condition $g_0 = a_F$ on the $R_0$ perturbation boundary ($g_0$ is the centripetal acceleration), as per the analysis of \S\ref{BTFR},  to give
\[
V_{flat}^4 = a_F\, G\, M_{flat},
\]
where $M_{flat}$ is the total estimated mass out to $V_{flat}$. The fact that this is an effect of the $\Sigma_F$ External Matter Field is manifested by the appearance of $a_F \equiv 4 \pi G \Sigma_F$.
\\\\
By contrast, in the MONDian analysis, the chosen form of the far-field force law \emph{ensures} that the BTFR, in Milgrom's form:
\[
V_{flat}^4 = a_0\, G\, M_{flat},
\]
arises automatically. In other words, the MOND far-field force law is directly emulating the $\Sigma_F$ External Matter Field.
\\\\
In summary, it is self-evident that the chosen form of MOND's far-field force law works as it does because it emulates the $\Sigma_F$ External Matter Field giving rise to a primary EFE in the form of the BTFR. We are left to conclude that \citet{Chae}  are in fact detecting secondary EFEs.
\subsection{Interim summary}
It has been shown that:
\begin{itemize}
	\item the characteristic acceleration $a_F \approx 1.2 \times 10^{-10}\,mtrs/sec^2$ is explicitly embedded in the SPARC data of \citet{Lelli2016A};
	\item equilibrium on the perturbation boundary requires $g_0 = a_F$;
	\item when $g_0 = a_F$, then the BTFR arises automatically from (\ref{eqn2}) but with MOND's acceleration scale $a_0$ replaced by $a_F$;
	\item the foregoing results lead to the unambiguous conclusion that $a_0$ and $a_F$ are one and the same thing;
	\item MOND works as it does because the far-field form of its force law is actually emulating the $\Sigma_F$ external matter field.
\end{itemize}
In summary, Milgrom's MOND is shown to be firmly anchored in the geometrical model of a Leibniz-Mach worldview.
\section{The inferred nature of the  unseen IGM} \label{IGM}
There are two basic questions to be answered here: 
\begin{itemize}
	\item  firstly, what is the distribution of material in the IGM?
	\item  secondly, what is the nature of this material?
\end{itemize}
We consider these question in turn.
\subsection{The distribution of material in the IGM} \label{IGM2}
The model (\ref{eqn2}) explicitly requires the existence of a non-trivial $D\approx 2$ quasi-fractal IGM of  unspecified material. In consequence, the quality \& precision of the results derived in \S\ref{aFcalc}, \S\ref{Gamma1} and \S\ref{BTFR} and appendix \S\ref{MassModel} stand as strong evidence in support of such an IGM.
\\\\
There is additional independent evidence: in a very recent paper \citet{Hong} construct the \lq{unseen matter}' map of the local universe (the local cosmic web) using a novel neural-network machine-learning algorithm. 
It is immediately striking from the images in this paper that this map shows all the voids, sheets and filaments that are such familiar features in the maps of galaxy distributions in the local universe, and the authors explicitly make this exact point. But, as  \citet{Baryshev1995} and many others have shown (see appendix \S\ref{Observations} for an overview), it is precisely these features  which characterize the  $D \approx 2$ quasi-fractal nature of galaxy distributions in the local universe. 
\\\\
Thus, although \citet{Hong} do not talk in terms of fractal distributions of unseen matter, the clear inference from their results, and the work of the astrophysical fractal community in general, is that their predicted distribution of unseen matter in the local cosmic web can be characterized as $D \approx 2$ quasi-fractal  - in other words, the idea of a $D \approx 2$ quasi-fractal unseen IGM is independently supported.
\subsection{The nature of the material in the IGM} \label{IGMmaterial}
Whilst the standard assumption is that the unseen IGM consists largely of the canonical Dark Matter of contemporary astrophysics, present considerations lead to a quite different conclusion: specifically, that rather than consisting largely of canonical Dark Matter, it must consist largely of near-perfect blackbody absorbers - such materials are no longer theoretical, they do exist courtesy of exotic carbon chemistry. See the discussion of \S\ref{IGM1}.
\\\\
The evidence of \S\ref{aFcalc}, \S\ref{Gamma1}, \S\ref{BTFR} and appendix \S\ref{MassModel} is unambiguous. For any given galactic object:
\begin{itemize}
	\item the absolute position of the perturbation boundary is located in the RC-fitting process (described in appendix \S\ref{MassModel}) and is labelled
	 $R_0^S$ when expressed in terms of SPARC's photometrically determined radial scalings; 
	\item the rotation velocity, $V_0$, on the perturbation boundary is therefore known;
	\item according to the theory of \S\ref{aFcalc} there is an absolute acceleration scale on the perturbation boundary represented by the invariant $\Gamma$ and estimated in \S\ref{Gamma1},  using SPARC rotation curves and photometry, to be  given by $\Gamma \approx 1.2 \times 10^{-10}\,mtrs/sec^2$;
	\item when the Galaxy Equilibrium Condition $g_0 = a_F$ (zero net radial acceleration on $R_0$) is used, then $g_0 = a_F = \Gamma$ arises automatically via (\ref{eqn2BB}) so that $V_0^2/R_0 \approx 1.2 \times 10^{-10}\,mtrs/sec^2$\,;
	\item given the prior information from MOND that $a_0 \approx 1.2 \times 10^{-10}\,mtrs/sec^2$, this latter result confirms that the absolute positions of the perturbation boundaries are correctly determined. In consequence, their radii $R_0$ can be estimated purely from disk dynamics;
	\item when we look at the plot of $(R_0,\,R_0^S)$ over the whole sample we find systematically:
	\begin{equation}
	0.1\,R_0^S < R_0 < R_0^S \label{eqn2GG}
	\end{equation}
\end{itemize}
where the distribution of $R_0/R_0^S$ is sharply modal with a peak at $0.25$.
Given that photometric methods of determining astrophysical distances are not detectably unreliable and that we expect $R_0 \sim R_0^S$, there is only one way that (\ref{eqn2GG}) can occur: there exists an unrecognized dimming mechanism which emulates \& therefore exaggerates the effect of the ordinary inverse-square dimming mechanism. 
\\\\
It is worth noting that if such an additional dimming mechanism does exist but is not recognized, then objects are automatically estimated to be further away than they really are, and therefore bigger than they really are, with the consequence that a missing-mass interpretation arises automatically. In such a case, it is easily seen that (\ref{eqn2GG}) if unrecognized would imply an equivalent missing mass equal to $96\%$ of the total mass budget which is entirely consistent with modern estimates of canonical Dark Matter in spiral disks.  But, given the evidence of (\ref{eqn2GG}) we are led to:\\\\
{\bf Hypothesis:} The material of the $D\approx 2$ quasi-fractal IGM consists largely of near-perfect blackbody absorbers since such a material distribution would give rise exactly to a systematic and \emph{undetectable} exaggeration of photometric distances, and hence of radial scales whilst, simultaneously, allowing a high degree of transparency in the IGM. Consequently, generally speaking, we would find $R_0 < R_0^S$.
\\\\
We expand upon the issues implicit to this hypothesis in \S\ref{IGM1} and \S\ref{Distances2}.
\subsection{Blackbody absorber material in the IGM ?}\label{IGM1}
Modern photometric methods of distance estimation are predicated upon one relevant implicit assumption in particular: that there is no such thing in nature as a perfect (or near-perfect) blackbody absorber. However, recent developments in the material sciences have shown how near-perfect blackbody absorbers can be created via exotic carbon chemistry in industrial processes for everyday usage. Consequently, the idea that such materials cannot exist in nature can no longer be sustained. 
\\\\
In particular, 
\citet{Mizuno}  were the first to show how to fabricate, from agglomerations of single-walled carbon nanotubes (SWCNTs), material distributions  having specific bulk statistics which act as near-perfect blackbody absorbers (emissivity $> 0.98$) across a very wide range of incident wavelengths from UV at 200$nm$ to the far IR at 200$\mu m$.  This behaviour has been shown to be independent of the specific properties of the individual SWCNTs, but is rather a consequence of the bulk statistical characteristics of the fabricated SWCNT distributions.
\\\\
We know that many allotropes of carbon exist in interstellar space and these must to some extent be blown into the IGM from the generality of galactic interiors. It is a short step to visualizing the existence of clouds of SWCNTs dispersed throughout the IGM  containing sub-populations which,  when viewed in projection along any given line of sight, possess the bulk statistical characteristics required to mimic the fabricated SWCNT distributions of \citet{Mizuno}. In this way, it is possible to conceive how SWCNT clouds within the IGM have the potential to act as \lq{dispersed near-perfect blackbody objects}'. 
\subsection{The implications of a quasi-fractal SWCNT cloud IGM} \label{Distances2}
Suppose that the $D \approx 2$ quasi-fractal IGM consists substantially of  \lq{dispersed near-perfect blackbody objects}' in the form of SWCNT clouds. What are the consequences arising?
\\\\ 
By virtue of its $D\approx 2$ quasi-fractal distribution, such an IGM would, to a significant extent, be transparent to radiation, which mirrors the primary reason why \citet{Charlier1908, Charlier1922, Charlier1924} suggested  the \lq{hierarchical universe}' as an early answer to the question \emph{Why is the sky dark at night?}
\\\\
But whilst a $D\approx 2$ quasi-fractal SWCNT IGM would, to a significant extent, be transparent to radiation,
it would by no means be totally transparent; broadly speaking and in addition to the usual inverse square dimming process, light from a source at distance $R$ would experience dimming via a process of near-perfect blackbody absorption by  SWCNT clouds in a way which would be proportional to $R^2$, again because this material is distributed quasi-fractally, $D\approx 2$. 
\\\\
This dimming mechanism would be indistinguishable in its effects from the ordinary inverse-square distance dimming process so that the total of observed dimming would be interpreted entirely as a distance effect. There are two consequences:
\begin{itemize}
	\item The principles underlying the process by which standard candles are used to estimate the absolute luminosities of distant objects are unchanged so that such estimates would not be affected by SWCNT cloud absorption, should the phenomenon actually exist;
	\item The photometric distance scale would be systematically exaggerated with the effect that objects of a given absolute luminosity would generally be estimated as being further away and larger than they actually are. Such an unrecognized exaggeration of linear scales taken at face value would automatically give rise to a \lq{missing mass}' problem within any given galaxy or, similarly, within any given galaxy cluster.
\end{itemize}
On this basis, it follows that for the SPARC data (for example) we can expect the photometric estimates of mass to remain unaffected and the length scales to be systematically exaggerated. 
\section{Conclusions}
When applying the geometrical Leibniz-Mach model to the galaxies of the SPARC sample there are, actually, two mutually exclusive possibilities for understanding the dynamical stability of the sample's disks:
\begin{enumerate}
	\item \emph{either assume that the photometrically determined radial scales (and hence distance scales) intrinsic to the SPARC sample can be taken face value} in which case, according to the model, the photometrically determined masses are typically around $5\sim10\%$ of theoretical requirements. In this case, we are led to conclude that there is a need for the canonical Dark Matter of modern astrophysics, both to bulk up individual galaxies but also to populate the IGM;
	\item \emph{or assume that the photometrically determined masses intrinsic to the SPARC sample can be taken at face value} in which case, according to the model, the photometrically determined radial scales (and hence distance scales) intrinsic to the SPARC sample are typically around $3\sim4\times$ greater than theoretical requirements. In this case, in order to account for the fact that photometric methods of determining distances are not detectably unreliable then, as per the arguments of \S\ref{IGMmaterial}, we are led to conclude that there is a need for near-perfect blackbody absorbers to populate the IGM.
\end{enumerate}
Whilst the canonical Dark Matter requirement (around $90\sim 95\%$ of the total mass) established by the first possibility for SPARC objects is perfectly consistent with Virial Theorem determinations, the primary analysis of this paper has established that it is the second possibility alone (the basis of the algorithm described in appendix \S\ref{Computation}) which resolves with quantitative precision the mysteries of MOND in the context of these objects.
\\\\
Accordingly, the second possibility must be favoured: the quasi-fractal IGM is populated with near-perfect blackbody absorbers which, by the arguments of \S\ref{Distances2}, if unrecognized leads to photometric methods significantly over-estimating the magnitudes of astrophysical length scales, thereby creating an automatic requirement for canonical Dark Matter to compensate.
\\\\
In conclusion, the choice is between:
\begin{itemize}
	\item either ubiquitous canonical Dark Matter, and no resolution of MOND's mysteries;
	\item or ubiquitous near-perfect blackbody absorbers and a quantitatively precise resolution of MOND's mysteries.
\end{itemize}
Finally, noting that galaxy clusters also have a Dark Matter requirement which is $90\% \sim 95\%$ of their total mass budgets, then the foregoing implies that the second possibility above is entirely sufficient to resolve the conundrums of galaxy clusters and their dynamics. 
\appendix
\section{A fractal universe: the observations \& the debate}\label{Observations}
A basic assumption of the \textit{Standard Model}
of modern cosmology is that, on some scale, the universe is homogeneous;
however, in early responses to suspicions that the accruing data was
more consistent with Charlier's conceptions \citet{Charlier1908, Charlier1922, Charlier1924}
of an hierarchical universe than with the requirements of the \textit{Standard
	Model},  \citet{De Vaucouleurs1970} showed that, within wide limits,
the available data satisfied a mass distribution law $M\approx r^{1.3}$,
whilst \citet{Peebles1980} found $M\approx r^{1.23}$. The situation,
from the point of view of the \textit{Standard Model}, continued to
deteriorate with the growth of the data-base to the point that, \citet{Baryshev1995}  were able to say
\begin{quote}
	\emph{...the scale of the largest inhomogeneities (discovered to date)
		is comparable with the extent of the surveys, so that the largest
		known structures are limited by the boundaries of the survey in which
		they are detected.}  
\end{quote}
For example, several redshift surveys of the late 20th century, such
as those performed by \citet{Huchra1983}, \citet{Giovanelli1986}, \citet{DeLapparent1988}, \citet{Broadhurst}, \citet{DaCosta} and \citet{Vettolani} 
etc discovered massive structures such as sheets, filaments, superclusters
and voids, and showed that large structures are common features of
the observable universe; the most significant conclusion drawn from
all of these surveys was that the scale of the largest inhomogeneities
observed in the samples was comparable with the spatial extent of
those surveys themselves.\\
\\
In the closing years of the century, several quantitative analyses
of both pencil-beam and wide-angle surveys of galaxy distributions
were performed: three examples are given by \citet{Joyce}  who analysed the CfA2-South catalogue to find fractal
behaviour with $D\,$=$\,1.9\pm0.1$; \citet{SylosLabini}
analysed the APM-Stromlo survey to find fractal behaviour with $D\,$=$\,2.1\pm0.1$,
whilst \citet{SylosLabini1} analysed the
Perseus-Pisces survey to find fractal behaviour with $D\,$=$\,2.0\pm0.1$.
There are many other papers of this nature, and of the same period,
in the literature all supporting the view that, out to $30-40h^{-1}Mpc$
at least, galaxy distributions appeared to be consistent with the simple stochastic fractal model with the critical fractal dimension of $D\approx  D_{crit} = 2$.\\
\\
This latter view became widely accepted (for example, see  \citet{Wu}), and the open question became whether or not
there was transition to homogeneity on some sufficiently large scale.
For example, \citet{Scaramella} analyse the ESO Slice
Project redshift survey, whilst \citet{Martinez} analyse
the Perseus-Pisces, the APM-Stromlo and the 1.2-Jy IRAS redshift surveys,
with both groups claiming to find evidence for a cross-over to homogeneity
at large scales.\\
\\
At around about this time, the argument reduced to a question of
statistics (\citet{Labini}, \citet{Gabrielli}, \citet{Pietronero}):
basically, the proponents of the fractal view began to argue that
the statistical tools (that is, two-point correlation function methods) widely used
to analyse galaxy distributions by the proponents of the opposite
view are deeply rooted in classical ideas of statistics and implicitly
assume that the distributions from which samples are drawn are homogeneous
in the first place.  \citet{Hogg}, having accepted
these arguments, applied the techniques argued for by the pro-fractal
community (which use the \emph{conditional density} as an appropriate
statistic) to a sample drawn from Release Four of the Sloan Digital
Sky Survey. They claimed that the application of these methods does
show a turnover to homogeneity at the largest scales thereby closing,
as they see it, the argument. In response, \citet{SylosLabini2} 
criticized their paper on the basis that the strength of the
conclusions drawn is unwarrented given the deficencies of the sample
- in effect, that it is not big enough. 
\\\\
More recently, \citet{Tekhanovich} have addressed the deficencies of the Hogg et al analysis by analysing the 2MRS catalogue, which provides redshifts of over 43,000 objects out to about 300Mpc, using conditional density methods; their analysis shows that the distribution of objects in the 2MRS catalogue is consistent with the simple stochastic fractal model with the critical fractal dimension of $D\approx  D_{crit} = 2$.
\\\\
To summarize, the proponents of non-trivially fractal large-scale
structure have won the argument out to medium distances and the controversy
now revolves around the largest scales encompassed by the SDSS.
\section{The geometric Leibniz-Mach model: outline} \label{LMC}
For all its familiarity, MOND remains a very odd construct, and it is this very oddness which alerts us to the idea that any underlying theory which provides for its fundamentals can be expected to deviate significantly from the canonical viewpoint. This point of view is considerably strengthened by the results of \citet{Chae}, which point to a breakdown of the Strong Equivalence Principle (SEP) which, if confirmed, poses a particular problem for General Relativity, thereby creating potential room for a significant shift in fundamental ideas.
\\\\
The geometric Leibniz-Mach model appeared in the mainstream literature in an early form as \citet{Roscoe2002}. Whilst this early form is only partially interpreted around the meaning of clocks and clock-rate synchronisation, the fully interpreted form exists in the archive as \citet{Roscoe2020}.
The following provides a brief outline.
\\\\
There are two issues to be considered: the nature of \emph{physical space} and the nature of \emph{physical time}.
\subsection{Leibniz briefly on the nature of physical space}
The debate of Clarke-Leibniz (1715$\sim$1716) (\citet{Alexander1984}) concerning the nature of physical space makes it clear that Leibniz considered the concept of the empty physical space to be a meaningless abstraction, and he held firmly to the view that the only significant thing was the set of relationships between \lq{objects}', whatever these \lq{objects}' might be.
\\\\
As a first step towards quantifying this idea in modern terms, we interpret it to imply the view that \emph{there is no such thing as a physical space which is empty} and \emph{metrical}. 
\\\\
We then formulate the question: 
\emph{how can one impose metric structure upon a physical space which is such that the metric structure becomes undefined when that physical space is empty?} 
\\\\
In order to provide a quantitative answer this question, it is instructive to reflect very briefly upon how we, as primitive human beings, form qualitative assessments of `distance' in our everyday lives without recourse to formal instruments.
\\\\
So, for example, when walking across a tree-dotted
landscape the changing \emph{angular} relationships between ourselves
and the trees provides the information required to assess both
\emph{what distance travelled?} and \emph{which tree nearer/further?} measured in units of human-to-tree angular displacements within that landscape. If we obliterate our view of the scene - say, with fog - then all forms of \lq{distance}' information are destroyed.
\\\\
In other words, the informal metric structure that we impose upon the space containing the landscape derives exclusively from changes in the angular relationships between ourselves and the elements of that landscape as we move within it. 
This indicates the geometric approach by which a formal metric structure can be imposed upon a specifically non-empty physical space.
\\\\
It is sufficient for present purposes to consider only the most simple possible case of an unbounded volume in which the total mass contained within any spherical surface does not vary. 
In particular, suppose that $\mathcal{M}(R)$ denotes the mass contained within an arbitrarily centred sphere of radius $R\equiv f(x^1,x^2,x^3)$ (for generalized curvilinear coordinates) and that this material-containing sphere  represents the non-empty space upon which a metric is to be imposed. Arguing from geometric first principles (\cite{Roscoe2002} or \cite{Roscoe2020}) we find that a metric structure for this physical three-space is projected out of its mass content according to:
\begin{eqnarray}
g_{ab} &=& \frac{1}{8 \pi \Sigma_F }\nabla_a \nabla_b \mathcal{M} \equiv \frac{1}{8 \pi \Sigma_F }\left(\frac{\partial^{2}\mathcal{M}}{\partial x^{a}\partial x^{b}}-\Gamma_{ab}^{k}\frac{\partial \mathcal{M}}{\partial x^{k}}\right),\label{3A}  \\
\Gamma_{ab}^{k} &\equiv&\frac{1}{2}g^{kj}\left(\frac{\partial g_{bj}}{\partial x^a}+\frac{\partial g_{ja}}{\partial x^b}-\frac{\partial g_{ab}}{\partial x^j} \right)\nonumber
\end{eqnarray} 
which represents a non-linear differential equation to be solved for $g_{ab}$ in terms of $\mathcal{M}(R)$. Here, $\Sigma_F$ is a parameter with dimensions of \emph{mass surface density} included to ensure that $g_{ab}$ is dimensionless,  and the factor $8\pi$ is included for later convenience. 
\\\\
This system can be resolved by making the modelling assumption that the coordinates $x^i$ can be Euclidean so that $R^2 = x^i x^j\,\delta_{ij}$. The metric tensor $g_{ab}$ is then explicitly determined to give the line element
\begin{equation}
ds^{2}\equiv g_{ij} dx^i dx^j \equiv \frac{1}{4 \pi\Sigma_{F}}\left[ \frac{ d_0 \mathcal{M} }{R^2}dx^{i}dx^{j}\delta_{ij}-\left(\frac{ d_0 \mathcal{M}}{R^2}-\frac{ \mathcal{M}' \mathcal{M}'}{4\, \mathcal{M}}\right) dR^2\right], \label{4F}
\end{equation}
which is invariant under rotations only. Here, $d_0$ is a dimensionless parameter of which more later.
Note that if $\mathcal{M}(R) \equiv 0$, then there is no metric space.
\subsection{Leibniz \& Mach briefly on the nature of time} \label{TimeDef}
Leibniz was equally clear in expressing his views about the nature of
time which are very similar to those expressed by \citet{Mach1960}.  They each viewed \emph{time} (specifically Newton's \emph{absolute time}) as a meaningless abstraction. In effect and in modern terms, for both \emph{time} is no more than a numerical label, $t$ say, which tracks sequential change within a physical system. 
\\\\
In the present context, the only kind of change which occurs is \emph{spatial displacement} so that, somehow or other, \emph{elapsed time} has to be quantified against the invariant displacement of (\ref{4F}). 
To this end,  (\ref{4F}) still allows the formal construction of the  variational principle
\[
\mathcal{I}(p,q)=\int^q_p{\cal L}\,dt\equiv \int^q_p\sqrt{g_{ij}\dot{x}^{i}\dot{x}^{j}}\,dt ,
\]
even though it remains obvious that minimizing $\mathcal{I}(p,q)$ with respect to  variations in $\mathbf{x}(t)$ cannot possibly lead to the determination of unique trajectories whilst the idea of \emph{elapsed time} remains unquantified. However, we find that this latter problem is automatically resolved by constraining the system to be such that \emph{energy is conserved}. This leads to two distinct cases: 
\subsubsection*{The general case of non-circular motions:} 
For this case, it is found that energy is only conserved if the elapsed time for a given displacement $d\mathbf{x}$ is \emph{defined} by
\begin{equation}
dt^2   \underset{\mathrm{def}}{=} \frac{1}{v_0^2}\left(\frac{\Sigma_F}{\Sigma_R}\right)^2\, g_{ij} dx^i dx^j 
\label{4G}
\end{equation}
where $v_0$ is an undetermined parameter with units of \emph{velocity} which calibrates clock rate, and $\Sigma_R$ is the mass surface density of the system at radius $R$ from the centre. 
\subsubsection*{The degenerate case of circular motions} 
For this case, energy is automatically conserved, and the elapsed time for a given displacement $R\,d\theta$ is found to be defined by
\begin{equation}
dt^2   \underset{\mathrm{def}}{=} \frac{1}{v_0^2}\left(\frac{\Sigma_R}{\Sigma_F}\right)\, R^2 d\theta^2. \label{3B}
\end{equation}
Equation (\ref{FirstEqn}), upon which this whole paper is predicated, arises directly from this degenerate state case. 
\subsection{The equilibrium state} \label{ESLC} 
When the mass distribution 
\begin{equation}
\mathcal{M}(R)=4\pi R^2\Sigma_F \label{3C}
\end{equation}
is constrained to be invariant under translations (that is, valid about any point), then it represents a $D=2$ fractal distribution of mass with a mass surface density $\Sigma_F$ which is independent of scale. In the following, it is shown that this constrained distribution is automatically in a state of dynamical equilibrium according to the Leibniz-Mach model.
\\\\
In practice, the unconstrained distribution (\ref{3C}) when substituted in (\ref{4F}) gives
\[
ds^2 \equiv g_{ij} dx^i dx^j = d_0\,\delta_{ij}\,dx^i dx^j + (1-d_0)\,dR^2 .
\]
Consequently, since the required distribution is constrained to be invariant under translations, then we must have $d_0=1$. Otherwise there is a centre, and the system is invariant under rotations only. 
\\\\
Since (\ref{3C}) gives $\Sigma_R = \Sigma_F$ as a trivial result, then (\ref{4G}) with $d_0=1$ gives directly
\begin{equation}
dt^2  = \frac{1}{v_0^2}\, \delta_{ij} dx^i dx^j \label{4H}
\end{equation}
as the definition of elapsed time for a displacement $d\mathbf{x}$ in a system characterized by a particular value for $v_0$. So, in an obvious sense, (\ref{4H}) acts as a clock, by which the passage of time can be determined. But $v_0$ is uninterpreted and so a question arises: given two distinct clocks of the (\ref{4H}) type, how can we ensure that these two clocks \emph{tick} at the same rate? 
\\\\
The answer is straightforward. Consider the purely classical point of view in which the passage of time is measured by an external classical clock. Suppose further that we consider the motion of an unaccelerated particle with a constant speed $v_0 = 10\,mtrs/sec$. Then we can either say that a displacement of $10\,mtrs$ occurs for every $1\,sec$ of elapsed time. Or we can say that $0.1\,sec$ elapses for every $1\,mtr$ of displacement. In the latter case, the moving particle is acting as a clock as per (\ref{4H}).
\\\\
It other words, assigning $v_0$ values in (\ref{4H}) according to the classical interpretation is equivalent to assigning $v_0$ values according to the condition that all Leibniz-Mach clocks are synchronized to \emph{tick} at the same rate and, automatically, at the same rate as an external classical clock.
\\\\
We now see that with this simple synchronization procedure, then the mass distribution $\mathcal{M}(R)=4\pi R^2\Sigma_F$, constrained to be valid about any point, represents a $D=2$ fractal distribution of mass which is automatically in a state of classical dynamical equilibrium.
\section{The recovery of $\Sigma_F$ from the SPARC sample} \label{MassModel}
\subsection{Object selection from the SPARC sample} \label{Gamma}
The SPARC sample compiled by \citet{Lelli2016A} consists of 175 nearby galaxies with modern surface photometry at $3.6\,\mu m$ and high quality rotation curves. The sample has been constructed to span very wide ranges in surface brightness, luminosity, rotation velocity and Hubble type, thereby forming a representative sample of galaxies in the nearby Universe. To date, the SPARC sample is the largest collection of galaxies with both high-quality rotation curves and NIR surface photometry.
\\\\
As we have noted, the galaxy model (\ref{eqn2G}) contains eight free parameters in all. Bearing this in mind,  then we select a subsample of SPARC objects according to the following criteria. Objects must have:
\begin{itemize}
	\item a rotation curve quality flag Q = 1 or 2;
	\item an inclination $> 30^o$;
	\item a SPARC estimate of $V_{flat}$ for the rotation curve. This is simply to ensure that the measured rotation curve is probably sufficiently sampled to allow a good fit of (\ref{eqn2G}) to it; 
	\item at least 10 rotation velocity measurements remaining after rejecting all those velocity measurements with a relative error $> 10 \%$.
\end{itemize}
These constraints whittle the original 175 SPARC objects down to a potentially usable sample of 90 objects. 
Having fitted (\ref{eqn2G}) to each of these 90 objects, we retain only those for which the rotation curves satisfy:
\begin{itemize}
	\item the condition that $R_0 < R_{max}$ where $R_{max}$ is the outermost radial measurement on the rotation curve. This is to ensure that $R_0$ is not obtained by extrapolating beyond the range of the rotation curve data;
	\item the condition that there are at least \emph{seven} data points on the interior region $0 < R \leq R_0$. The reasoning here is simply that of the eight model parameters, $(V_{flat}, R_0, M_0, a_0 ... a_4)$, seven are directly associated with quantities that are either on or interior to the perturbation boundary $R=R_0$ and should therefore be determined by interior measurements.  
\end{itemize}
Depending on how $\Upsilon_* \in (0.5,1.0)$ is chosen, this gives a final sample of between 60 and 70 objects which are used to determine the invariant $\Gamma$ of (\ref{eqn2BB}) in the manner described in \S\ref{Computation}.
\subsection{Numerical model}
The primary objective is the recovery of $\Sigma_F$ (and hence of $a_F$) from SPARC rotation curves and photometry using the model (\ref{eqn2}) and, to this end, we need the best possible estimates for the three parameters $(R_0, M_0, V_{flat})$ for each object. 
\\\\
There is now a choice: do we use SPARC photometry with mass model (\ref{MassModel1}) to represent $\mathcal{M}_g(R) \equiv M_0\,F(R/R_0)$ in (\ref{eqn2}) directly (the MOND approach in practice), or do we  replace $\mathcal{M}_g(R) \equiv M_0\,F(R/R_0)$ by a highly parameterized empirical mass-model, and then use the parameter space of this model to help optimize the rotation curve fit for each object which, by definition, will then optimize the estimates of  $(R_0, M_0, V_{flat})$?
\\\\
We choose the empirical mass modelling route and so, writing $F(R/R_0) \approx f(R/R_0,\mathbf{a})$ where $\mathbf{a}$ represents the five-dimensional parameter space of the  $f$-model defined at (\ref{MM1}), then (\ref{eqn2}) becomes
\begin{eqnarray}
\left[\frac{V_{rot}(R)}{V_{flat}}\right]^2 \Sigma_R &=& \Sigma_F, ~~~~ R < \infty; \label{eqn2G} \\  
\Sigma_R &\equiv& \frac{M_0\,f(R/R_0,\mathbf{a})}{4 \pi R^2},~~~~ R \leq R_0; \nonumber \\
\Sigma_R &\equiv& \frac{M_0  + 4 \pi (R^2-R_0^2)\,\Sigma_F}{4 \pi R^2},~~~~ R > R_0, \nonumber
\end{eqnarray}
which is now the basis for the computation. The rotation curve fit for each object is optimized with respect to variation of the eight parameters $(R_0, M_0, V_{flat},\mathbf{a})$.
\subsection{The $\Sigma_F$ recovery algorithm} \label{Computation}
The computation follows the theory of \S\ref{aFcalc}. Whilst the process is straightforward, $M_0$, $R_0$ and $\Sigma_F$ have to be rescaled at various stages according to the symmetries of that theory. Consequently, the notation needs careful attention:
\begin{itemize}
	\item $\Sigma_F^*$ is the unknown value of the mass surface density parameter, to be calculated;
	\item $R_0^S$ is the radius of the perturbation boundary expressed in terms of SPARC's radial scales; 
	\item $M_0^S$ is the mass contained within $R_0^S$ estimated using SPARC photometry and the mass model of \citet{Lelli2016B} given at (\ref{MassModel1});
	\item $R_0$ and $M_0$ are interim rescalings of $R_0^S$ and $M_0^S$ respectively;
	\item $R_0^*$ is the final rescaling of $R_0^S$ whilst $M_0^* \equiv M_0^S$ always;
	\item $V_0^*$ is the rotation velocity at $R_0^*$ (or, equivalently, at $R_0^S$ or $R_0$) computed from (\ref{eqn2}).
\end{itemize}
A basic assumption of the algorithm is that the total estimated mass within $R_0^S$ for each object must
be determined using SPARC photometry \& the mass model (\ref{MassModel1}). In practice, this means that at the end of the rotation curve fitting process for each object, the total computed mass, $M_0$, must  be rescaled using symmetry (\ref{eqn2A}) so that $M_0 \rightarrow M_0^S \equiv M_0^*$. With this understanding, the $\Sigma_F$ recovery algorithm proceeds as follows:
\begin{enumerate}
	\item The value of the invariant $\Gamma$ of (\ref{eqn2BB}) is independent of any value assigned to $\Sigma_F$. So, in the first instance, set $\Sigma_F = \Sigma_F^0$ where $\Sigma_F^0$ is some trial value. In practice, any $\Sigma_F^0 \sim 1.0$ will do;
	\item Then, for each SPARC object, optimize the fit of (\ref{eqn2G}) to the rotation curve by varying the eight local parameters $(V_{flat},R_0^S,M_0, \mathbf{a})$ where $\mathbf{a}\equiv (a_0...a_4)$ are the parameters of the empirical mass-model (\ref{MM1}). Since $\Sigma_F^0$ is a trial value, symmetry (\ref{eqn2C}) shows that the computed $M_0$ for any given object cannot generally coincide with its  photometric counterpart, $M_0^S$, which is initially unknown since $R_0^S$ is also initially unknown; 
	\item Once the value of $R_0^S$ for each SPARC object has been estimated via the optimization process, we can use SPARC photometry with mass-model (\ref{MassModel1}) to estimate $M_0^S$;
	\item Using the symmetry (\ref{eqn2A}), we now map $M_0 \rightarrow M_0^S$ for each object according to
	\begin{equation}
	M_0 \rightarrow \beta M_0 = M_0^S \equiv M_0^*,~~~~ R_0^S \rightarrow \beta^{1/2} R_0^S \equiv R_0~~{\rm where}~~\beta = \left(\frac{M_0^S}{M_0}\right).  \label{eqn2H}
	\end{equation}
	As required, all masses in the $\Sigma_F^0$ trial solution are now calibrated against SPARC photometry. The radial scaling has also necessarily changed; 
	\item Symmetry (\ref{eqn2B}) maps $\Sigma^0_F \rightarrow \Sigma^*_F$ and also has $\Gamma \equiv g_0^2/a_F$ of (\ref{eqn2BB}) as an invariant. Consequently, $\Gamma$ can be evaluated from either of the $\Sigma^0_F$ or $\Sigma^*_F$ solutions. But the $\Sigma^*_F$ solution remains unknown, and so we evaluate $\Gamma$ from the $\Sigma^0_F$ solution;
	\item For the evaluation: using (\ref{eqn2H}), the $\Sigma^0_F$ solution gives both  $a_F = 4 \pi G \Sigma_F^0$ and $g_0 = {V_0^*}^2/R_0$ so that the invariant $\Gamma \equiv g_0^2/a_F$  becomes known; 
	\item Once $\Gamma$ is known, we can use the equilibrium condition of (\ref{eqn2CC}), $g_0^*\equiv {V_0^*}^2/R_0^* = a_F^*$, to evaluate the required final state corresponding to $\Sigma_F^*$ using $g_0^* = a_F^* =\Gamma$ from (\ref{eqn2DD}).
\end{enumerate}
\section{Some computational details}\label{MassModels}
There are various details which are necessary to reliably reproduce the results of this paper.
\subsection*{The empirically defined mass distribution}
In practice, it is found that the five-parameter model
\begin{eqnarray}
f(R/R_0,\mathbf{a}) &\equiv& Y^{a_0}, \label{MM1}
\\
Y & \equiv & \left[(1-a_1-a_2-a_3-a_4)\,X+a_1\,X^3+a_2\,X^5+a_3\,X^7+a_4\,X^9\right], \nonumber \\
X &\equiv& \frac{R}{R_0}, \nonumber \\
\mathbf{a} &\equiv& \left(a_0, a_1, a_2, a_3, a_4\right) \nonumber
\end{eqnarray}
provides for an extremely accurate fit of the theory to the SPARC rotation curves and hence, by inference, provides a high fidelity modelling of the mass distributions.
\subsection*{The minimization metric}
For minimization problems involving noisy data, it is generally considered best to use a metric based on the $L_1$-norm. So, for a rotation curve with measured velocities $V_{sparc}(R_i),\,i=1..N$, we seek to determine the disk parameters $(R_0, M_0, V_{flat})$ by minimizing:
\[
metric = \sum^N_{i=1} \left| \frac{V_{sparc}(R_i)-V_{theory}(R_i)}{V_{sparc}(R_i)} \right|
\]
with respect to variation in them. This gives far superior results to those arising from use of the $L_2$-norm, for example.
\subsection*{The Nelder-Mead iteration}
Because the data is very noisy, we estimate the eight free parameters  ($R_0, V_{flat}, M_0,\mathbf{a}$), where $\mathbf{a} \equiv (a_0, a_1, a_2, a_3, a_4)$, as follows:
\begin{enumerate} 
	\item Choose a set of randomly generated initial guess for each of the free parameters;
	\item Run the Nelder-Mead minimization process for, typically, about 25000 ($2.5\times 10^4$) times per disk before the results completely settle down. For most objects, this is far more than is necessary, but there are a few awkward objects.
\end{enumerate}
\section{Least-areas linear regression} \label{LeastAreas}
In the following, the quantity being minimised is independent of how the predictor/response pair is chosen for the regression. The result is a linear model which can be algebraically inverted to give the exact linear model which would also arise from regressing on the interchanged predictor/response pair. So, any inference drawn about the relationship between the predictor and response is independent of how the predictor/response pair is chosen.
\\\\
Imagine a straight line drawn through a two-dimensional scatter diagram. Every point in this diagram subtends a right-angled triangle onto the straight line such that the two short sides of the triangle are parallel to the coordinate axes and meet at the point concerned. The least-areas linear regression is then defined to be the particular line which minimizes the total area of all the triangles summed over all the points. 
\\\\ 
Suppose that we have the data $(X_i, Y_i),~i = 1..N$ to which we fit the model
\[
y = A x + B.
\]
It is a simple matter to show that the least-areas regression simply requires that:
\[
A = \pm \sqrt{\frac{\left(\Sigma Y_i\right)^2-N \,\Sigma Y_i^2}{\left(\Sigma X_i\right)^2-N\, \Sigma X_i^2} }, ~~~
B = \frac{\left( \Sigma Y_i - A\, \Sigma X_i\right)}{N}.
\]
To demonstrate the algebraically invertable property, fit the model $x = \alpha y + \beta$ to the same data, to get:
\[
\alpha = \pm \sqrt{\frac{\left( \Sigma X_i\right)^2 - N \, \Sigma X_i^2}{\left( \Sigma Y_i\right)^2 - N \, \Sigma Y_i^2}}, ~~~~ \beta = \frac{\Sigma X_i - \alpha \, \Sigma Y_i}{N}.
\]
Comparing the two models quickly shows that $\alpha = 1/A$ and $\beta = -B/A$ so that $x=\alpha y + \beta$ is the algebraic inverse of $y = A X + B$.
\newpage
\subsection*{Data availablity statement}
The data underlying this article were provided by Stacy McGaugh (Stacy.McGaugh@case.edu) under licence / by permission. Data will be shared on request to the corresponding author with permission of Stacy McGaugh.

\end{document}